\documentclass{article}
\usepackage[utf8]{inputenc}

\title{Atomic Swaptions: Cryptocurrency Derivatives}
\author{James A. Liu  \\Department of Computer Science\\ University of California, Irvine\\ \texttt{jamesal1@uci.edu}}

\usepackage{tikz}
\usetikzlibrary{shapes.geometric, arrows,positioning}
\usetikzlibrary{shapes,arrows,fit,calc,positioning,automata}

\tikzstyle{aona} = [rectangle, minimum width=3.2cm, minimum height=1cm, draw=red, fill=red!20,thick,align=left]
\tikzstyle{bonb} = [rectangle, minimum width=3.2cm, minimum height=1cm,text centered, draw=blue, fill=blue!20,thick,align=left]
\tikzstyle{aonb} = [rectangle, minimum width=3.2cm, minimum height=1cm,text centered, draw=red,fill=blue!20, thick,align=left]
\tikzstyle{bona} = [rectangle, minimum width=3.2cm, minimum height=1cm,text centered, draw=blue,fill=red!20, thick,align=left]
\tikzstyle{process} = [rectangle, minimum width=3cm, minimum height=1cm, text centered, draw=black, fill=orange!30]
\tikzstyle{decision} = [rounded rectangle, minimum width=2.5cm, minimum height=1cm, text centered, draw=black,thick,align=center]

\tikzstyle{arrow} = [thick,->,>=stealth]
\tikzset{fontscale/.style = {font=\relsize{#1}}}
\tikzset{XOR/.style={draw,circle,thick,append after command={
        ([yshift=-.05cm]\tikzlastnode.north) edge ([yshift=.05cm]\tikzlastnode.south)
        ([xshift=-.05cm]\tikzlastnode.east) edge ([xshift=.05cm]\tikzlastnode.west)
        }
    }
}
\tikzset{every picture/.style={/utils/exec={\sffamily}}}
\tikzstyle{every node}=[font=\scriptsize,style={thick}]
\tikzstyle{every edge}=[style={draw=black,thick}]

\usepackage{circuitikz}
\usepackage[numbers]{natbib}
\usepackage{graphicx}
\usepackage{environ}
\usepackage{caption}
\usepackage{subcaption}
\usepackage{float}
\usepackage{url}

\makeatletter
\newsavebox{\measure@tikzpicture}
\NewEnviron{scaletikzpicturetowidth}[1]{%
  \def\tikz@width{#1}%
  \begin{lrbox}{\measure@tikzpicture}%
  \BODY
  \end{lrbox}%
  \pgfmathparse{#1/\wd\measure@tikzpicture}%
  \BODY
}
\makeatother

\begin{document}

\maketitle
\begin{abstract}
The atomic swap protocol allows for the exchange of cryptocurrencies on different blockchains without the need to trust a third party. However, market participants who desire to hold derivative assets such as options or futures would also benefit from trustless exchange.\\
In this paper we propose the atomic swaption, which extends the atomic swap to allow for trustless exchange of option contracts. Crucially, atomic swaptions do not require the use of oracles. We also introduce the margin contract, which provides the ability to create leveraged and short positions. Lastly, we discuss how atomic swaptions may be routed on the Lightning Network.
\end{abstract}
\section{Trustless Exchange}
Blockchains such as Bitcoin rely on cryptographic protocols to secure ownership of digital assets. However, as originally designed, there was no mechanism for different blockchains to communicate securely with each other. Thus, cryptocurrency trading is done "off-chain", or outside of the scope of the blockchains involved. Counterparties who wish to trade online typically deposit their funds into a trusted third party, typically taking the form of a centralized cryptocurrency exchange.

The need to trust centralized exchanges has long been an uncomfortable reality within the cryptocurrency community. Exchange participants often fear that their funds may be frozen, seized or stolen; history has shown that such fears are well-founded\cite{hack2018}. Conversely, exchanges with good reputations have monopolistic power over potential new users/investors of a given cryptocurrency, allowing them to extract rents and exercise undue influence over cryptocurrency development in general. For instance, exchanges may charge large transaction fees. They may choose which cryptocurrencies can be traded on their exchange, perhaps charging a listing fee. Creators of new cryptocurrencies must cede to their demands or take their chances by working with less reputable exchanges. These frictions make it difficult for participants to acquire coins, posing an impediment to the adoption of cryptocurrencies.

The volatility of cryptocurrencies provides another impediment to their adoption, gamblers and speculators notwithstanding. Potential users might be more inclined to adopt a given cryptocurrency if they could hedge against volatility risk through the use of financial options or futures markets. For example, producers of cryptocurrencies (i.e. miners) have revenues denominated in cryptocurrency, but generally have expenses denominated in various fiat currencies. They might want to use futures contracts to cover this mismatch.

However, relying on centralized exchanges for derivatives trades would expose users to the same problems that spot traders face. In fact, these problems would be magnified. The long-term nature of derivatives means that users would have to keep their assets on the exchange for the entire length of the contract. In addition, any leverage offered by the exchange would lead to direct and indirect counterparty risk; the exchange may enter into one-sided bets or be lax in enforcing margin for some of its users. In the case of extreme market fluctuations, the exchange may become insolvent.

The atomic swap allows users to avoid the risks and disadvantages inherent to centralized exchanges when trading. Atomic swaps use Hashed Time Lock Contracts (HTLCs), a form of cryptographic escrow, with multiple blockchains acting as the trusted third party. Viewed another way, atomic swaps utilize cryptographic secrets to allow users to prove on one blockchain that a prerequisite payment occurred on a different blockchain.

In this paper we demonstrate that the functionality of atomic swaps can be further extended to allow the creation of option contracts, which we term atomic swaptions. In particular, parties can deposit only a fractional margin instead of the full principal. This, in turn, allows for a meaningful notion of a futures contract, enabling users to take leveraged or short positions on one cryptocurrency against another.

\subsection{Related Work}
Related approaches to decentralized cryptocurrency derivatives focus on trading ERC20 tokens on the Ethereum blockchain \cite{zx} \cite{feas2018} and/or use oracles (trusted providers of external data) to determine contract outcomes \cite{dlc}. In contrast, the approach we outline can be used between different blockchains, with only the requirement that they can support the creation of HTLCs with the same hash function. Namely, the blockchains do not need to be Turing-complete (e.g. Ethereum). In addition, atomic swaptions do not require oracles, eliminating an unnecessary element of trust and interaction from the process.

Others have also modified atomic swaps to create basic option contracts \cite{frei2013} \cite{fern2018}. However, the construction they use has a serious drawback: if either party reneges, the participants cannot get refunded until option expiration. The approach described in this paper follows a principle of using a larger number of simpler transactions to implement the desired functionality. This leads to greater flexibility in adding helpful features to the contract.

\section{Layout}
In this paper, we review how atomic swaps work, then show how atomic swaptions naturally arise as an extension of them. We show how various features can be implemented, such as early cancellation and margin. Furthermore, we discuss how atomic swaptions could be used in practice. In particular, we discuss how they may be implemented on the Lightning Network, as well as the general ramifications of having long-term contracts on the Lightning Network. Finally, we note some of the risks and limitations of atomic swaptions.\\
In the example transactions we use, Alice initially possesses ACoin and Bob initially possesses BCoin, placeholder cryptocurrencies that live on different blockchains with functionality equivalent to Bitcoin.\\
The diagrams are styled after the Lightning Network white paper\cite{lightning2016}, with some differences. Boxes correspond to transactions; the capsules provide a stylistic interpretation of the act of publishing a transaction. Red fill color indicates the ACoin blockchain, and blue fill color indicates the BCoin blockchain. A red border indicates that Alice can publish the transaction, and a blue border indicates that Bob can publish the transaction. The vertical order of the boxes loosely correspond to the temporal sequence of events; a box directly below another represents a transaction that should be published immediately in response to the other. The numerical values should not be taken literally; in particular, 1 timestep should be interpreted as a quantity that allows parties to comfortably respond to their counterparty's actions in the appropriate manner.
\newpage
\section{Hashed Timelock Contract}
\subsection{Motivation}
Hashed Timelock Contracts\cite{lightning2016} are a way to prove receipt of coins. If Bob wants to prove payment, it is not sufficient for him to simply send coins to an address given by Alice, as there is no feasible way to prove to a different blockchain that such a transaction successfully occurred. Instead, Bob sends coins to an HTLC.
\subsection{Construction}
 An HTLC is a contract that has two alternate redemption conditions. In this context, a transaction fulfilling the first condition is called a "payment transaction". It requires that Alice signs it and that she provides the pre-image of a hashed value, or a "secret", that only she knows. This is the so-called "hashlock". A transaction fulfilling the second condition is called a "refund transaction". It requires Bob's signature and that a certain amount of time has elapsed. This is the so-called "timelock".\\ This provides the desired functionality: In order to take the coins, Alice must have her secret published on the blockchain, and if she doesn't, then Bob can take back the coins once the timelock has expired.
 
\begin{figure}[H]
\noindent\makebox[\textwidth]{
\resizebox{\textwidth}{!}{
\begin{tikzpicture}[node distance=.5cm]
\begin{scope}[blend mode=multiply]

\node (startb) [bonb,yshift=1cm] {Funding Tx\\
Bob broadcasts\\
Output: HTLC 1 BCoin};
\node (startx) [XOR,scale=2, below = of startb]{};

\node (arev) [decision,left = of startx, xshift = -1cm] {Alice accepts:\\
reveals A};
\node (anop) [decision,right = of startx, xshift = 1cm] {Alice silent:\\
reveals nothing};

\node (exb) [aonb, below = of arev]
    {Swap Tx \\
    Only Alice can broadcast
    \vspace{.5em} \\
    Secret: A
    \vspace{.5em} \\
    Output: Alice 1 BCoin};

\node (reb) [bonb, below = of anop]
    {Refund Tx \\
    Only Bob can broadcast
    \vspace{.5em} \\
    T Locktime
    \vspace{.5em} \\
    Output: Bob 1 BCoin};

\draw [->,draw = black,thick]
(startb) edge (startx)
(startx.west) edge (arev)
(arev) edge (exb)
(anop) edge (reb)
(startx) -- (anop)
{};
\end{scope}
\end{tikzpicture}
}
}
\caption{Hashed Timelock Contract.}
\end{figure}
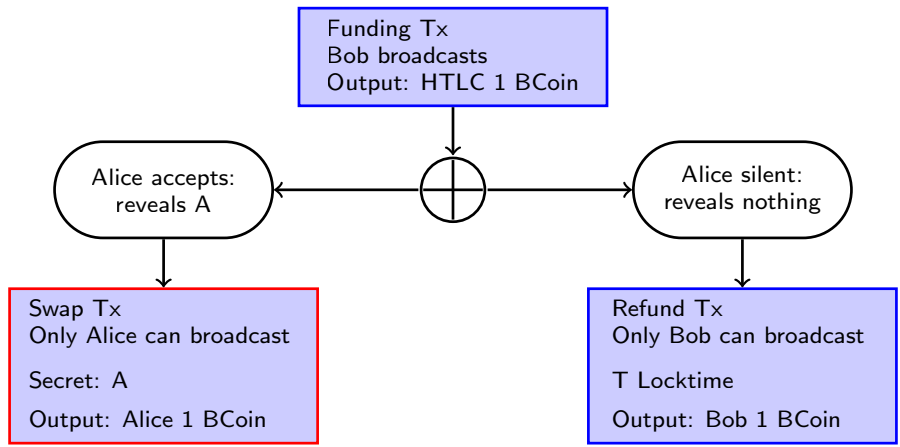
 
\newpage
\section{Atomic Swap} \label{swap}
\subsection{Motivation}
Suppose Alice wants to trade her ACoin for Bob's BCoin. If she were to send her Acoin to Bob first, Bob could then simply drop out of the process without sending any BCoin to her, and vice versa if Bob sends his BCoin first. Instead, Alice and Bob can use an atomic swap, which eliminates the possibility of theft.
\subsection{Construction}
 An atomic swap\cite{nolan} consists of two HTLCs, one on each blockchain. First, Alice generates a secret $A$, and sends ACoin to an HTLC with hash $H(A)$, expiring at time $T+1$. Once the ACoin has been sent to the HTLC, Bob sends BCoin to an HTLC with hash $H(A)$, expiring at time $T$. If Alice publishes the transaction to take the BCoin, then in doing so she reveals $A$. Bob can then use $A$ to take the ACoin, since the first HTLC expires later than the second. This completes the swap.\\
If a party drops out of the process before Alice reveals $A$, each party can recover their funds through the respective HTLC's refund transaction.
\begin{figure}[H]
\noindent\makebox[\textwidth]{
\resizebox{\textwidth}{!}{
\begin{tikzpicture}[node distance=.5cm]
\begin{scope}[blend mode=multiply]

\node (starta) [aona] {Funding Tx\\
Alice broadcasts\\
Output: HTLC 1 ACoin
};
\node (startb) [bonb, below right = of starta,yshift=1cm] {Funding Tx\\
Bob broadcasts\\
Output: HTLC 1 BCoin};
\node (startx) [XOR,scale=2, below = of startb, xshift = -.9cm]{};

\node (arev) [decision,left = of startx, xshift = -1cm] {Alice accepts:\\
reveals A};
\node (anop) [decision,right = of startx, xshift = 1cm] {Alice reneges:\\
reveals nothing};

\node (exb) [aonb, below = of arev]
    {Swap Tx \\
    Only Alice can broadcast
    \vspace{.5em} \\
    Secret: A
    \vspace{.5em} \\
    Output: Alice 1 BCoin};

\node (exa) [bona, below = of exb,yshift = .5cm]
    {Swap Tx \\
    Only Bob can broadcast
    \vspace{.5em} \\
    Secret: A
    \vspace{.5em} \\
    Output: Bob 1 ACoin};
    
\node (reb) [bonb, below = of anop]
    {Refund Tx \\
    Only Bob can broadcast
    \vspace{.5em} \\
    T Locktime
    \vspace{.5em} \\
    Output: Bob 1 BCoin};
    
\node (rea) [aona, below = of reb,yshift = .5cm]
    {Refund Tx \\
    Only Alice can broadcast
    \vspace{.5em} \\
    T+1 Locktime
    \vspace{.5em} \\
    Output: Alice 1 ACoin};

\draw [->,to path={-| (\tikztotarget)}, draw = black,thick]
(starta) edge (startx.north)
(startb) edge (startx.north)
{};
\draw [->,draw = black,thick]
(startx.west) edge (arev)
(arev) edge (exb)
(anop) edge (reb)
(startx) -- (anop)
{};
\end{scope}
\end{tikzpicture}
}
}
\caption{Atomic swap, composed of two HTLCs. Under normal operation, both Swap Tx's or both Refund Tx's will be successful.}
\end{figure}
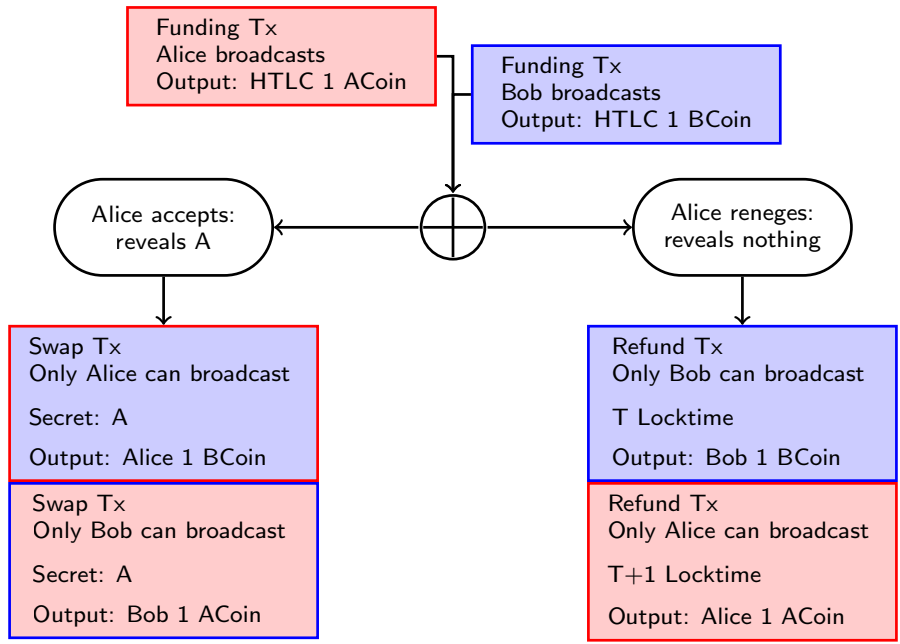

\section{Atomic Swaption} \label{swaption}
\subsection{Motivation}
Suppose now that Alice wants to purchase from Bob a financial option to trade ACoin for BCoin. A financial option is a contract - stylistically, a piece of paper signed by Bob - that says Alice can choose to trade 1 ACoin for 1 BCoin with Bob up until time $E$. If, before time $E$, Alice wants to make this trade, she can present this contract to Bob, along with 1 ACoin, and Bob is obligated to give her 1 BCoin in return. She can also let the contract expire by doing nothing. The option contract is valuable for Alice because she can wait to see how ACoin and BCoin change in price relative to each other; at the expiration of the option contract, Alice will usually take the more valuable of 1 ACoin or 1 BCoin, and Bob will have the less valuable one. In return, Bob receives a payment of ACoin from Alice when he writes the contract, known as a premium.\\
Simply put, to do an option trade, Alice pays Bob a premium, and Bob gives the contract to Alice.
\subsection{Observation}
In the earlier atomic swap example, notice that once Bob has funded the BCoin HTLC, Alice can choose to go through with the trade or to wait until the timelocks expire and refund herself. In other words, the atomic swap functions as an option for Alice.\\
This is generally seen as a shortcoming of the atomic swap, since it may incentivize Alice to negotiate swaps in bad faith. There has been some discussion of this issue \cite{fairness} and how to resolve it \cite{harmful}\cite{comit}; in this paper, we make the simplifying assumption that $T$ can be set to be a short period of time such that Alice's "option value" is negligible, but still long enough for Alice and Bob to properly execute the protocol.
\subsection{Construction}
The atomic swaption consists of two nested atomic swaps. The first atomic  swap ("funding contract") acts as the option trade; it pays Bob a premium while simultaneously funding the second atomic swap, which acts as the option contract. The second atomic swap looks identical to the original atomic swap, except that the expiration $E$ is set to be a large value, say, 1 year. More succinctly, an atomic swaption is an atomic swap that swaps a premium for the creation of an option (which happens to be implemented as another atomic swap).\\
Note that the principal deposit transaction of the HTLCs in the funding contracts should require both Alice and Bob to sign them, in addition to revealing the secret. Before funding these contracts, Alice signs the ACoin principal deposit transaction and Bob signs the BCoin principal transaction, so that the other party cannot change the output(s) of the transaction.

\begin{figure}[H]
\noindent\makebox[\textwidth]{
\resizebox{\textwidth}{!}{
\begin{tikzpicture}[node distance=.5cm]
\begin{scope}[blend mode=multiply]
\node (starta) [aona] {Funding Tx\\
\vspace{.5em} \\
Alice broadcasts\\
\vspace{.5em} \\
Output:\\ Funding Contract 1.1 ACoin
};
\node (startb) [bonb, below right = of starta,yshift=1cm] {Funding Tx\\
\vspace{.5em} \\
Bob broadcasts\\
\vspace{.5em} \\
Output:\\ Funding Contract 1 BCoin};
\node (startx) [XOR,scale=2, below = of startb, xshift = -.9cm]{};

\node (arev) [decision,left = of startx, xshift = -1cm] {Alice accepts:\\
reveals A};
\node (anop) [decision,right = of startx, xshift = 1cm] {Alice reneges:\\
reveals nothing};

\node (trb) [aonb, below = of arev]
    {Principal Deposit Tx \\
    Only Alice can broadcast
    \vspace{.5em} \\
    Secret: A
    \vspace{.5em} \\
    Output:\\
    \hspace*{.9em} Swaption Contract 1 BCoin };

\node (tra) [bona, below = of trb,yshift = .5cm]
    {Principal Deposit Tx \\
    Only Bob can broadcast
    \vspace{.5em} \\
    Secret: A
    \vspace{.5em} \\
    Outputs:\\
    0. Bob 0.1 ACoin \\
    1. Swaption Contract 1 ACoin};
    
\node (reb) [bonb, below = of anop]
    {Refund Tx \\
    Only Bob can broadcast
    \vspace{.5em} \\
    T Locktime
    \vspace{.5em} \\
    Output: Bob 1 BCoin};
    
\node (rea) [aona, below = of reb,yshift = .5cm]
    {Refund Tx \\
    Only Alice can broadcast
    \vspace{.5em} \\
    T+1 Locktime
    \vspace{.5em} \\
    Output: Alice 1.1 ACoin};

\node (opx) [XOR,scale=2, below = of startx, yshift = -2.5cm]{};

\node (a2rev) [decision,left = of opx, xshift = -1cm] {Alice exercises:\\
reveals A2};
\node (a2nop) [decision,right = of opx, xshift = 1cm] {Alice lets expire:\\
reveals nothing};

\node (exb) [aonb, below = of a2rev]
    {Exercise Tx \\
    Only Alice can broadcast
    \vspace{.5em} \\
    Secret: A2
    \vspace{.5em} \\
    Output: Alice 1 BCoin};

\node (exa) [bona, below = of exb,yshift = .5cm]
    {Exercise Tx \\
    Only Bob can broadcast
    \vspace{.5em} \\
    Secret: A2
    \vspace{.5em} \\
    Output: Bob 1 ACoin};
    
\node (expb) [bonb, below = of a2nop]
    {Expiration Tx \\
    Only Bob can broadcast
    \vspace{.5em} \\
    E Locktime
    \vspace{.5em} \\
    Output: Bob 1 BCoin};
    
\node (expa) [aona, below = of expb,yshift = .5cm]
    {Expiration Tx \\
    Only Alice can broadcast
    \vspace{.5em} \\
    E+1 Locktime
    \vspace{.5em} \\
    Output: Alice 1 ACoin};

\draw [->,to path={-| (\tikztotarget)}, draw = black,thick]
(starta) edge (startx.north)
(startb) edge (startx.north)
(tra) edge (opx)
(trb) edge (opx)
{};
\draw [->,draw = black,thick]
(startx.west) edge (arev)
(arev) edge (trb)
(anop) edge (reb)
(startx) edge (anop)
(a2rev) edge (exb)
(a2nop) edge (expb)
(opx) edge (a2nop)
(opx.west) -- (a2rev)
{};

\end{scope}
\end{tikzpicture}
}
}
\caption{Atomic swaption, composed of two atomic swaps.}
\end{figure}
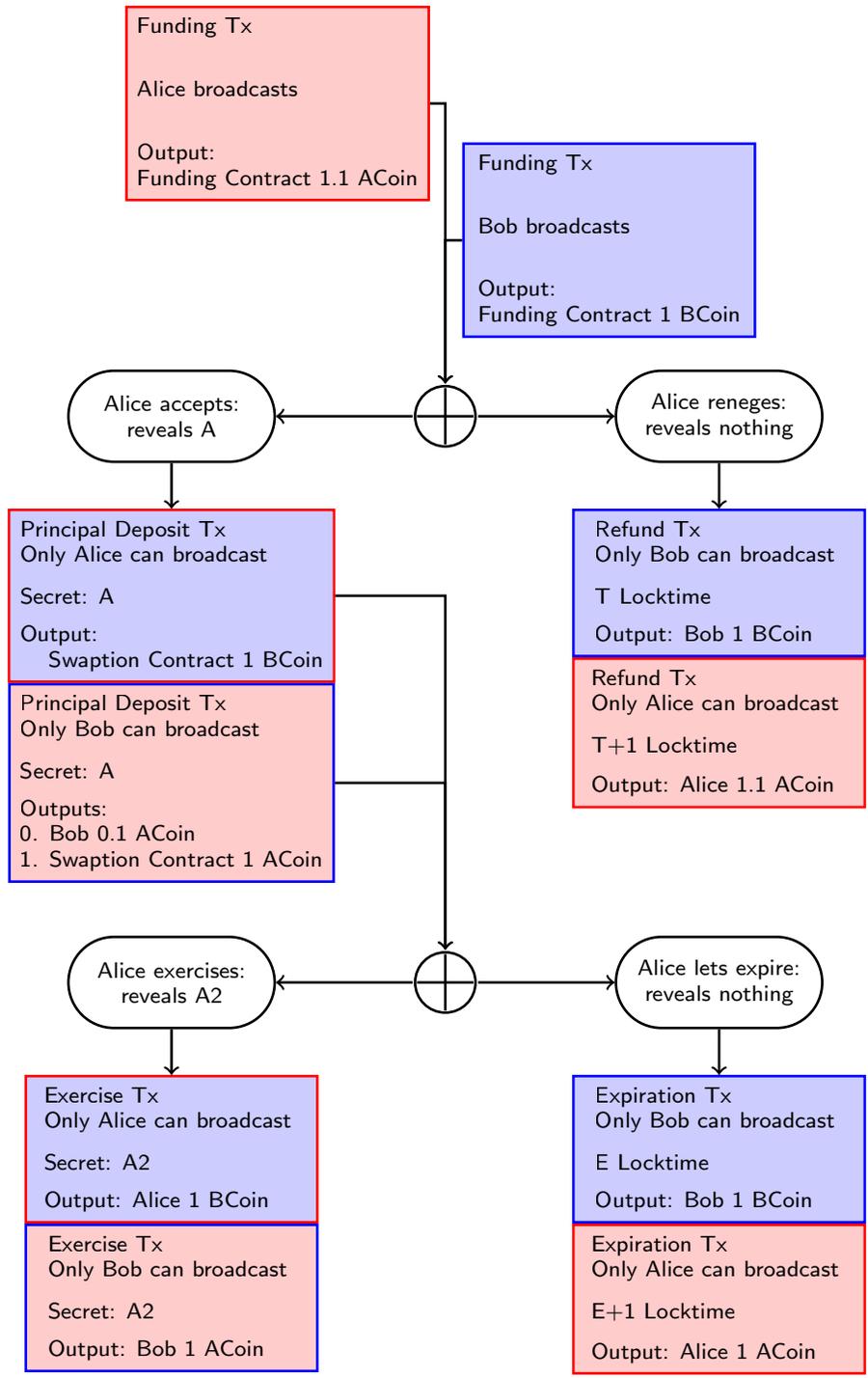

\section{Extensions of the Atomic Swaption}
\subsection{Early Cancellation}
\subsubsection{Motivation}
Suppose Alice would like to be able to forfeit the swaption and reclaim her ACoin early, and be able to do so without cooperation with Bob.
\subsubsection{Construction}
 The ACoin swaption expiration transaction is changed to require Alice to provide a different secret $A3$ to take back her ACoin, instead of a timelock; the BCoin swaption expiration transaction allows Bob to take back his coins immediately after by providing $A3$, or after expiration $E$. However, this creates a race condition, where Alice may try to claim both the coins. To prevent this, the funds are sent to a further set of HTLCs. The contract corresponding to Alice's exercise transaction allows Bob to take the BCoin if Alice also published the cancel transaction, and vice versa. Thus, if Alice tries to cheat, Bob can take all of the funds as punishment.

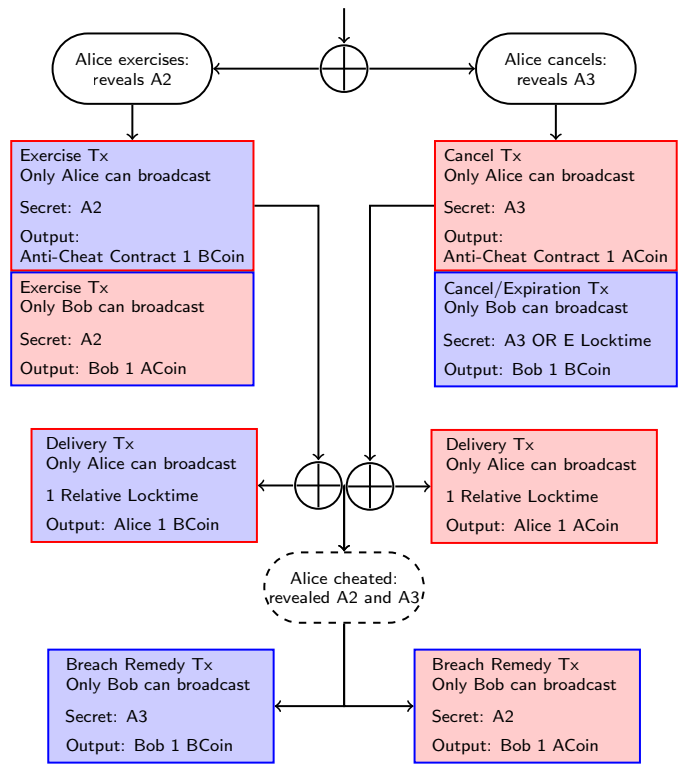
\begin{figure}[H]
\noindent\makebox[\textwidth]{
\resizebox{.75\textwidth}{!}{
\begin{tikzpicture}[node distance=.5cm]
\begin{scope}[blend mode=multiply]
]

\node (opx) [XOR,scale=2]{};

\node (funding) [coordinate,above = of opx] {};

\node (a2rev) [decision,left = of opx, xshift = -1cm] {Alice exercises:\\
reveals A2};

\node (exb) [aonb, below = of a2rev]
    {Exercise Tx \\
    Only Alice can broadcast
    \vspace{.5em} \\
    Secret: A2
    \vspace{.5em} \\
    Output:\\ Anti-Cheat Contract 1 BCoin};

\node (exa) [bona, below = of exb,yshift = .5cm]
    {Exercise Tx \\
    Only Bob can broadcast
    \vspace{.5em} \\
    Secret: A2
    \vspace{.5em} \\
    Output: Bob 1 ACoin\hspace*{3.15em}};

\node (acan) [decision,right = of opx, xshift=1cm] {Alice cancels:\\
reveals A3};

\node (cana) [aona, below = of acan]
    {Cancel Tx \\
    Only Alice can broadcast
    \vspace{.5em} \\
    Secret: A3
    \vspace{.5em} \\
    Output:\\ Anti-Cheat Contract 1 ACoin};

\node (canb) [bonb, below = of cana,yshift=.5cm]
    {Cancel/Expiration Tx \\
    Only Bob can broadcast
    \vspace{.5em} \\
    Secret: A3 OR E Locktime
    \vspace{.5em} \\
    Output: Bob 1 BCoin\hspace*{3.1em}};

\node (exx) [XOR,scale=2, below right= of exa,xshift= .15cm,yshift=-.4cm]{};
\node (canx)[XOR,scale=2, below left = of canb,xshift=-.15cm,yshift=-.4cm]{};

\node (acheat) [decision,below = of opx,yshift=-6cm,dashed] {
Alice cheated:\\
revealed A2 and A3};

\node (cheata) [bonb, below left = of acheat] {
    Breach Remedy Tx\\
    Only Bob can broadcast
    \vspace{.5em} \\
    Secret: A3
    \vspace{.5em} \\
    Output: Bob 1 BCoin};

\node (cheatb) [bona, below right= of acheat]
    {Breach Remedy Tx \\
    Only Bob can broadcast
    \vspace{.5em} \\
    Secret: A2
    \vspace{.5em} \\
    Output: Bob 1 ACoin};
    
\node (relb) [aonb, left = of exx] {
    Delivery Tx\\
    Only Alice can broadcast
    \vspace{.5em} \\
    1 Relative Locktime
    \vspace{.5em} \\
    Output: Alice 1 BCoin};
    
\node (rela) [aona, right = of canx] {
    Delivery Tx\\
    Only Alice can broadcast
    \vspace{.5em} \\
    1 Relative Locktime
    \vspace{.5em} \\
    Output: Alice 1 ACoin};

\draw [->,to path={-| (\tikztotarget)}, draw = black,thick]
(exb) edge (exx)
(cana) edge (canx)
(exx) edge (acheat)
(canx) edge (acheat)
;

\draw [->,to path={|- (\tikztotarget)}, draw = black,thick]
(opx.east) -- (acan)
(acheat) edge (cheata)
(acheat) edge (cheatb)
;
\draw [->,draw = black,thick]
(funding) edge (opx)
(a2rev) edge (exb)
(opx.west) -- (a2rev)
(acan) edge (cana)
(exx) edge (relb)
(canx) edge (rela)
;

\end{scope}
\end{tikzpicture}
}
}
\caption{Early Cancellation. Unchanged portions have been omitted.}
\end{figure}

\subsection{Margin}
\subsubsection{Motivation}
A downside of the vanilla atomic swaption is that all of the funds involved ("principal") must be locked up in the contract until exercise or cancellation. In traditional financial markets, Alice would not need to provide any funds ahead of time, and Bob would only have to deposit some minimal amount of collateral, or "margin", to secure his option position.

\subsubsection{Construction}
Instead of depositing all of their funds into the contract, Alice and Bob deposit an agreed-upon margin into so-called margin contracts. Bob's margin contract requires both Alice and Bob to sign; Alice signs a transaction that allows Bob to send the margin to the swaption contract, but it specifies that the output is equal to the full principal, thus requiring that he also deposit the rest of the funds as well. Alice uses SIGHASH\textunderscore ANYONECANPAY so that Bob can choose where to deposit the funds from. Bob signs a transaction with timelock $M$ that sends the margin to Alice; this acts as the margin expiration.\\
If Bob does not deposit funds into the swaption contract by the margin expiration, Alice can take the margin funds, and does not need to reveal any secret to do so, meaning that she can also take back her ACoin by not exercising. Alice's margin contract functions similarly to Bob's, but it expires before Bob's margin contract does. If she fails to deposit the principal, then Bob can take her margin, and forfeit his own.

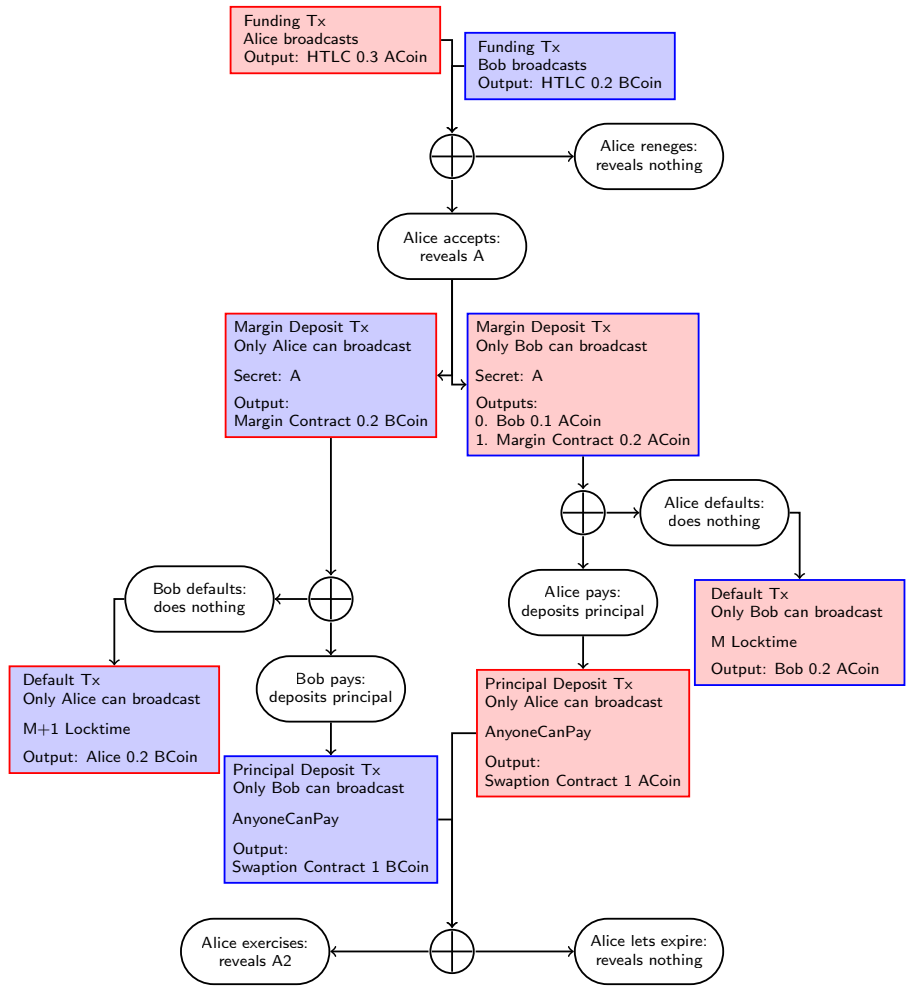
\begin{figure}[H]
\noindent\makebox[\textwidth]{
\resizebox{\textwidth}{!}{
\begin{tikzpicture}[node distance=.5cm]
\begin{scope}[blend mode=multiply]

\node (starta) [aona] {Funding Tx\\
Alice broadcasts\\
Output: HTLC 0.3 ACoin
};
\node (startb) [bonb, below right = of starta,yshift=1cm] {Funding Tx\\
Bob broadcasts\\
Output: HTLC 0.2 BCoin};
\node (startx) [XOR,scale=2, below = of startb, xshift = -.9cm]{};

\node (arev) [decision,below = of startx] {Alice accepts:\\
reveals A};
\node (anop) [decision,right = of startx, xshift = 1cm] {Alice reneges:\\
reveals nothing};

\node (trb) [aonb, below = of arev,xshift=-1.85cm]
    {Margin Deposit Tx \\
    Only Alice can broadcast
    \vspace{.5em} \\
    Secret: A
    \vspace{.5em} \\
    Output:\\ Margin Contract 0.2 BCoin};

\node (tra) [bona, below = of arev,xshift=2cm]
    {Margin Deposit Tx \\
    Only Bob can broadcast
    \vspace{.5em} \\
    Secret: A
    \vspace{.5em} \\
    Outputs:\\
    0. Bob 0.1 ACoin \\
    1. Margin Contract 0.2 ACoin};

\node (bmarx) [XOR,scale=2, below = of tra]{};

\node (adef) [decision, right = of bmarx]
{Alice defaults:\\
does nothing};

\node (defa) [bona, below = of adef,xshift=1.3cm]
    {Default Tx \\
    Only Bob can broadcast
    \vspace{.5em} \\
    M Locktime
    \vspace{.5em} \\
    Output: Bob 0.2 ACoin};

\node (apay) [decision, below = of bmarx]
{Alice pays:\\
deposits principal
};

\node (paya) [aona, below = of apay]
    {Principal Deposit Tx \\
    Only Alice can broadcast
    \vspace{.5em} \\
    AnyoneCanPay
    \vspace{.5em} \\
    Output:\\
    Swaption Contract 1 ACoin};
    
\node (amarx) [XOR,scale=2, below = of trb,yshift=-.8cm]{};

\node (bdef) [decision, left = of amarx]
{Bob defaults:\\
does nothing};

\node (defb) [aonb, below = of bdef,xshift=-1.3cm]
    {Default Tx \\
    Only Alice can broadcast
    \vspace{.5em} \\
    M+1 Locktime
    \vspace{.5em} \\
    Output: Alice 0.2 BCoin};

\node (bpay) [decision, below = of amarx]
{Bob pays:\\
deposits principal
};

\node (payb) [bonb, below = of bpay]
    {Principal Deposit Tx \\
    Only Bob can broadcast
    \vspace{.5em} \\
    AnyoneCanPay
    \vspace{.5em} \\
    Output:\\
    Swaption Contract 1 BCoin};

\node (opx) [XOR,scale=2, below = of arev, yshift = -4.7cm]{};

\node (a2rev) [decision,left = of opx, xshift = -1cm] {Alice exercises:\\
reveals A2};
\node (a2nop) [decision,right = of opx, xshift = 1cm] {Alice lets expire:\\
reveals nothing};

\draw [->,to path={|- (\tikztotarget)}, draw = black,thick]
(arev) edge (trb.east)
(arev) edge (tra.west)
;

\draw [->,to path={-| (\tikztotarget)}, draw = black,thick]
(starta) edge (startx.north)
(startb) edge (startx.north)
(adef) edge (defa)
(bdef) edge (defb)
(paya) edge (opx)
(payb) edge (opx)
;
\draw [->,draw = black,thick]
(startx) edge (arev)
(startx) edge (anop)
(opx) edge (a2nop)
(opx.west) -- (a2rev)
(trb) edge (amarx)
(tra) edge (bmarx)
(amarx) edge (bpay)
(amarx) edge (bdef)
(bmarx) edge (apay)
(bmarx) edge (adef)
(bpay) edge (payb)
(apay) edge (paya)
;

\end{scope}
\end{tikzpicture}
}
}
\caption{Swaption with margin. Unchanged portions have been omitted.}
\end{figure}

\subsubsection{Strategic default}
When negotiating terms, Alice and Bob should take into account the possibility of strategic default, which changes the payoff function of the swaption. Setting Alice's margin-to-principal ratio to be equal to Bob's is sufficient to ensure that Alice never has an incentive to default; however, Bob will have an incentive to default if ACoin drops enough in value relative to BCoin. In light of this, we outline two possible approaches to designing margin. The fixed margin approach is essentially "set and forget"; once the swaption is funded, Alice and Bob do not take any action until the swaption is about to expire. The floating margin approach tries to closely replicate the traditional payoff function of a call or put option, but requires frequent interaction between Alice and Bob.

\subsubsection{Fixed margin swaption}
\begin{figure}[H]
    \centering
     \begin{subfigure}{\textwidth}
      \centering
    \includegraphics[scale=.4]{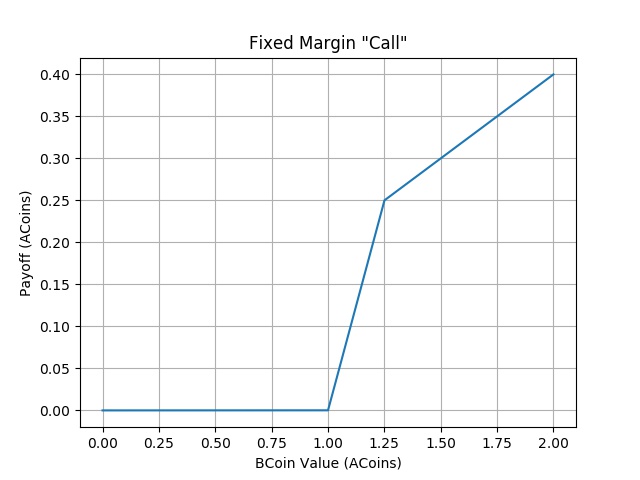}
    \end{subfigure}
    \begin{subfigure}{\textwidth}
     \centering
    \includegraphics[scale=.4]{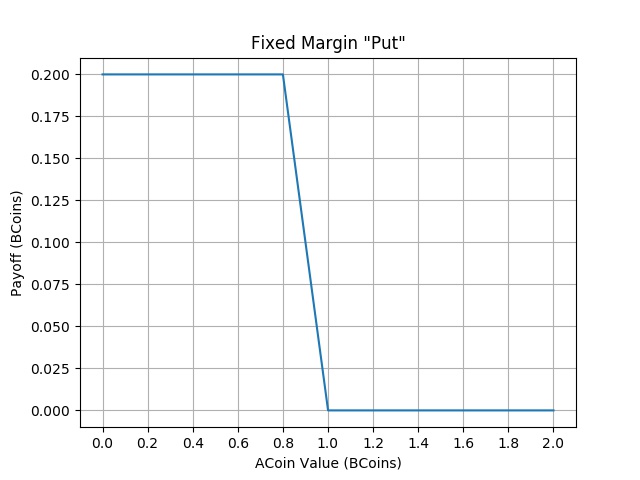}
    \end{subfigure}
    \caption{Fixed margin swaption payoff functions. Note that both graphs describe the same swaption. When denominated in ACoin it is a call option; when denominated in BCoin, it is a put option.}
    \label{fig:mainfig}
\end{figure}

Here, the margin expiration $M$ is set immediately before the swaption expiration ($E-2$ at the latest), so that Bob decides at expiration whether or not to default. Let $m_B,p_B$ correspond to the amount of BCoin margin and principal, and likewise for $p_A$. The intrinsic value of the swaption is then given by $max(0,min(m_B,p_B-p_A))$, not including the value of the margin deposit. This leads to the kinked payoff function shown in Figure \ref{fig:mainfig}; it can be decomposed into two traditional options for the purposes of pricing.

\subsubsection{Floating margin swaption}
Here, Alice and Bob can agree to continuously "mark to market" the option, increasing or decreasing the amount of margin required such that the expected gain from strategic default is below a small, pre-determined threshold. This is done by atomically cancelling the old swaption while creating the new swaption. The margin expiration is set to an intermediate value $M$, perhaps ranging from a day to a week. If Alice and Bob cooperate, the time until the margin expiration will continuously be reset to this value, but if one party stops cooperating, then the two parties have time to come up with the funds to avoid defaulting. The choice of $M$ is a tradeoff; a smaller value of $M$ allows for smaller margin and/or a lower probability of strategic default, but provides less time to respond in the non-cooperative case.\\
Floating margin is best done on the Lightning Network; otherwise, it could be very expensive to repeatedly create and cancel swaption contracts using on-chain transactions.

\subsubsection{Futures}
Since floating margin accurately replicates the payoff of a call/put option, it can be used for the creation of futures contracts. A version of put-call parity states that a long futures contract is equivalent to a long call option and short put option position (disregarding early exercise). In other words, to create a futures contract, Alice and Bob can atomically open two floating margin swaptions simultaneously with the same strike price and expiration, allowing Alice to swap 1 ACoin for 1 BCoin and allowing Bob to swap 1 BCoin for 1 ACoin. Note that this construction does not cancel out; Alice contributes ACoin to both contracts, and Bob contributes BCoin to both contracts. The strike price should be kept close to the futures market price (more correctly, the expected spot price at expiration), in order to minimize the amount of margin needed.

\subsubsection{Limitations}
Margin contracts introduce a form of transaction malleability in that the transaction ID of the principal deposit transaction is dependent on the source of the remaining principal, which is unknown ahead of time. The above construction is not affected by this issue, but more complex constructions may require transactions downstream of margin contracts to be signed ahead of time. These would require the use of SIGHASH\textunderscore NOINPUT \cite{bip118}, which is not currently included in Bitcoin.\\
It should be noted that Alice's margin deposit is only necessary due to the limitations of HTLCs as a communication protocol between blockchains. That is, Bob cannot prove on the BCoin blockchain that Alice \textit{didn't} deposit funds into the ACoin swaption contract (and Alice cannot prove on the BCoin blockchain that she \textit{did} deposit the ACoin).\\
If this were possible, then Alice's exercise transaction, which occurs on the BCoin blockchain, could be prevented until she deposited her ACoin into the ACoin swaption contract. If, however, ACoin and BCoin were on the same blockchain (e.g. they were both ERC20 tokens), then the contract could be designed so that Alice would not need to deposit any funds until she decides to exercise. 

\section{Financial State Channels}
Here we use swaptions as an example for how long term contracts may perform on the Lightning Network; the following techniques and analysis may be useful for a variety of other contracts such as discreet log contracts.
\subsection{Swaptions on the Lightning Network}
If Alice and Bob want to perform an atomic swap on the Lightning Network \cite{lnswap}, they can do so by constructing a chain of Hashed Time Lock Contracts that sends ACoin along a path of intermediaries from Alice to Bob, and then a chain of HTLCs that sends BCoin along a path of intemediaries from Bob to Alice. The intermediaries simply have to mimic the action of their successive neighbor, and thus Alice and Bob are virtually connected.\\
Similarly, to do an atomic swaption, they can open swaption funding contracts along the same path. Moreover, each payment channel can be updated to send funds directly to the most recent state of the swaption; this avoids the need to play out the full series of transactions if the channel is closed.\\
Since the channel's state can send to an arbitrary number of outputs, long-running contracts do not impair operation of the channels, beyond the specific amount of funds that have to be tied up in the contract.\\
Within a payment channel, users only need enough funds to cover the margin to participate in a swaption contract. However, they may have to close the channel in order to pay the principal if they do not have enough funds in the channel. 

\subsection{Third Party Decoupling and Unwinding}
If a third party, Carol, has nodes appearing on both the ACoin and BCoin paths between Alice and Bob, she can instead decouple the swaption, using her own secret with Bob. In this case, Carol functions as counterparty to both Alice and Bob (or to other intermediaries acting similarly), and has zero net position. This lowers the required time for swaption exercise (since each contract on a channel must expire 1 timestep later than its downstream neighbor's contract). Furthermore, Carol can unilaterally close her long option position with Bob without waiting for Alice to do so, freeing up her funds sooner. In the fully cooperative case, it also simplifies the option unwinding process.\\
Suppose now that an identical trade passes through Alice and Carol such that Carol would open a long position against Alice. Since Alice and Carol would now have a net zero position against each other, it would be ideal to end the process such that both swaptions have been cancelled, so that their funds are not locked up needlessly. Carol can set her swaption with Alice to expire 1 timestep after Alice’s swaption with her, using the same secrets. Alice would then reveal her secret, causing a circular flow of funds that can be undone with cyclic rebalancing\cite{cyclic}.\\
Similarly, by using self-trading, existing positions can be rerouted, such as when two parties want to close their channel without putting any contracts onto the blockchain. Use of these techniques can help minimize the costs of keeping swaption contracts open on the Lightning Network.

\tikzstyle{c} = [circle,draw]
\begin{figure}[H]
    \centering
     \begin{subfigure}{\textwidth}
        \centering
        \begin{tikzpicture}
          [scale=1]
          \node (nd)[c] at (0,0) {Dave};
          \node (na)[c] at (2,0) {Alice};
          \node (nb)[c] at (6,0) {Bob};
          \node (nc)[c] at (4,0) {Carol};
        
          \foreach \from/\to/\sec in {na/nc/A,nc/nb/C}{
            \draw[->,red]    (\from) to[out=30,in=150] node[below=-.4cm]{\sec} (\to);
            \draw[<-,blue]   (\from) to[out=-30,in=-150] node[below=-.4cm]{\sec} (\to);
            }
        \end{tikzpicture}
        \caption{Bob sells a swaption to Alice.}
    \end{subfigure}
    \begin{subfigure}{\textwidth}
        \centering
        \begin{tikzpicture}
          [scale=1]
          \node (nd)[c] at (0,0) {Dave};
          \node (na)[c] at (2,0) {Alice};
          \node (nb)[c] at (6,0) {Bob};
          \node (nc)[c] at (4,0) {Carol};
        
          \foreach \from/\to/\sec in {na/nc/A,nc/nb/C}{
            \draw[->,red]    (\from) to[out=30,in=150] node[below=-.4cm]{\sec} (\to);
            \draw[<-,blue]   (\from) to[out=-30,in=-150] node[below=-.4cm]{\sec} (\to);
            }
            \draw[<-,red]    (na) to[out=60,in=120] node[below=-.4cm]{A} (nc);
            \draw[->,blue]    (na) to[out=-60,in=-120] node[below=-.4cm]{A} (nc);
            \draw[<-,red]    (nd) to[out=30,in=150] node[below=-.4cm]{A'} (na);
            \draw[->,blue]    (nd) to[out=-30,in=-150] node[below=-.4cm]{A'} (na);
        \end{tikzpicture}
        \caption{Dave sells an identical swaption to Carol.}
    \end{subfigure}
    \begin{subfigure}{\textwidth}
        \centering
        \begin{tikzpicture}
          [scale=1]
          \node (nd)[c] at (0,0) {Dave};
          \node (na)[c] at (2,0) {Alice};
          \node (nb)[c] at (6,0) {Bob};
          \node (nc)[c] at (4,0) {Carol};
        
            \draw[->,red]    (nc) to[out=30,in=150] node[below=-.4cm]{C} (nb);
            \draw[<-,blue]   (nc) to[out=-30,in=-150] node[below=-.4cm]{C} (nb);
            \draw[<-,red]    (nd) to[out=30,in=150] node[below=-.4cm]{A'} (na);
            \draw[->,blue]    (nd) to[out=-30,in=-150] node[below=-.4cm]{A'} (na);
        \end{tikzpicture}
        \caption{Alice and Carol cancel their swaptions.}
    \end{subfigure}
\label{fig:nodes}
\end{figure}

\subsection{Economic Considerations}

Long paths are likely to be expensive, as Alice and Bob will have to compensate all of the intermediaries for the time value of their locked-up funds, as well as for taking on the risks and possible network transaction fees inherent in the process. Note also that intermediaries will have to maintain sufficient liquidity in order to avoid defaulting on margin swaptions, and need to be compensated for this. Swaptions, DLCs, and other long-term contracts all compete for the same network fund capacity.\\
This suggests that Lightning Network-based derivatives trading will mostly be fairly centralized, with a small set of densely connected large traders/exchanges, and many smaller traders directly connected to the large traders.\\
The ability to unwind identical swaptions encourages standardization of contract terms such as strike price, size, expiry, etc.

\section{Practical Considerations}
As with other long-term contracts, there is a risk that there is an emergency rule change on one of the blockchains, such as in the hashing or signing algorithm. Presumably, coins not moved within a certain timeframe will be burned (rendered unspendable). There are a few possibilities for Alice and Bob; the dynamics are influenced by the particular contract and the timing of the rule change. They may cooperate and reopen the contract under the new rules, possibly with one party extorting the other for an excess payment. If they do not cooperate, Alice must choose between exercise or cancelling by the move deadline. Therefore, Alice loses some of the time/"extrinsic" value of the swaption. However, Alice can also harm Bob by waiting until the compromised coins are burned, then taking the coins from the other blockchain.\\
Leveraged trading is unlikely to be user-friendly, as it may require a good deal of familiarity with option pricing formulas/software. In addition, users of margin should not rely on their counterparty being cooperative, as the counterparty often stands to gain substantially if margin is forfeited. Given a substantial enough swaption market, and an excessive usage of leverage among participants, it is possible that manipulators will attempt to orchestrate "short squeezes" by accumulating large swaption positions whose margin expirations occur at the same time, thus creating a scramble for funds and/or forfeiture of margin. They may be aided through use of long paths of intermediaries on the Lightning Network, as this increases the number of participants who need to obtain funds.\\
It is recommended that timelocks be set to use timestamps rather than block height, so that the timelocks expire in the intended order; it is difficult to predict the relative ordering of blocks far in the future between two different blockchains.\\
There have been a variety of attempts to create cryptocurrencies pegged to fiat currencies, also known as stablecoins. Users seeking to minimize volatility will likely prefer to use stablecoins in their swaption trades, though this of course requires them to trust in the underlying peg.

\section{Conclusion}
The design and adoption of new blockchains is an ongoing process, and it remains an open question what operations a blockchain should support to best facilitate the development of applications on top of it. We have demonstrated that a wide class of inter-blockchain derivatives can be implemented with only the features required to implement an atomic swap. The atomic swaption thus provides a practical path to derivatives trading on existing cryptocurrencies, and removes the need for designers of future blockchains to create specialized derivatives applications.

\bibliographystyle{unsrt}
\raggedright \bibliography{references}

\begin{thebibliography}{10}

\bibitem{hack2018}
Andrea Tan and Yuji Nakamura.
\newblock Cryptocurrency markets are juicy targets for hackers: Timeline.
\newblock
  \url{https://www.bloomberg.com/news/articles/2018-01-29/cryptocurrency-markets-are-juicy-targets-for-hackers-timeline}.
\newblock Accessed: 2018-07-10.

\bibitem{zx}
Antonio Juliano.
\newblock dydx: A standard for decentralized derivatives.
\newblock April 24, 2018.

\bibitem{feas2018}
Shayan Eskandari, Jeremy Clark, Vignesh Sundaresan, and Moe Adham.
\newblock On the feasibility of decentralized derivatives markets.
\newblock {\em CoRR}, abs/1802.04915, 2018.

\bibitem{dlc}
Thaddeus Dryja.
\newblock {\em Discreet Log Contracts}.
\newblock MIT Digital Currency Initiative.

\bibitem{frei2013}
Jorge Timón.
\newblock Freimarkets.
\newblock \url{https://github.com/jtimon/freimarkets}, 2013.
\newblock Accessed: 2018-07-24.

\bibitem{fern2018}
Fernando Nieto.
\newblock Trust-minimized derivatives.
\newblock
  \url{https://gist.github.com/fernandonm/75cf0b0381ed92404e8a651dd790f75d},
  2018.
\newblock Accessed: 2018-07-24.

\bibitem{lightning2016}
Joseph Poon and Thaddeus Dryja.
\newblock The bitcoin lightning network: Scalable off-chain instant payments.
\newblock January 14, 2016.

\bibitem{nolan}
TierNolan.
\newblock Alt chains and atomic transfers.
\newblock
  \url{https://bitcointalk.org/index.php?topic=193281.msg2224949#msg2224949}.
\newblock Accessed: 2019-11-07.

\bibitem{fairness}
Runchao Han, Haoyu Lin, and Jiangshan Yu.
\newblock On the optionality and fairness of atomic swaps.
\newblock In {\em Proceedings of the 1st ACM Conference on Advances in
  Financial Technologies}, AFT '19, pages 62--75, New York, NY, USA, 2019. ACM.

\bibitem{harmful}
Dan Robinson.
\newblock Htlcs considered harmful.
\newblock
  \url{http://diyhpl.us/wiki/transcripts/stanford-blockchain-conference/2019/htlcs-considered-harmful/}.
\newblock Accessed: 2019-11-07.

\bibitem{comit}
Lloyd~Fournier Thomas~Eizinger and Phillip Hoenisch.
\newblock The state of atomic swaps.
\newblock \url{https://scalingbitcoin.org/transcript/tokyo2018/atomic-swaps}.
\newblock Accessed: 2019-11-07.

\bibitem{bip118}
Christian Decker.
\newblock Bip118.
\newblock \url{https://github.com/bitcoin/bips/blob/master/bip-0118.mediawiki},
  2018.
\newblock Accessed: 2018-10-17.

\bibitem{lnswap}
Aaron van Wirdum.
\newblock Atomic swaps: How the lightning network extends to altcoins.
\newblock
  \url{https://bitcoinmagazine.com/articles/atomic-swaps-how-the-lightning-network-extends-to-altcoins-1484157052}.
\newblock Accessed: 2019-11-07.

\bibitem{cyclic}
Sebastián Reca.
\newblock Rebalancing in the lightning network: Analysis and implications.
\newblock
  \url{http://diyhpl.us/wiki/transcripts/scalingbitcoin/tokyo-2018/rebalancing-lightning/}.
\newblock Accessed: 2019-11-07.

\end{thebibliography}

\pagebreak


\end{document}